# The microwave induced resistance response of a high mobility 2DEG from the quasi-classical limit to the quantum Hall regime

**S. A. Studenikin,[a,*] M. Byszewski,[b] D. K. Maude,[b] M. Potemski,[b] A. Sachrajda,[a] Z. R. Wasilewski, M. Hilke,[c] L. N. Pfeiffer,[d] K. W. West[d]**

[a]*Institute for Microstructural Sciences, NRC, Ottawa, Ontario K1A-0R6, Canada*

[b]*Grenoble High Magnetic Field Laboratory, MPI/FKF and CNRS, Grenoble 38-042, France*

[c]*Department of Physics, McGill University, Montreal H3A 2T8, Canada*

*Bell Laboratories, Lucent Technologies, Murray Hill, New Jersey* 07974-0636*, USA*

---

**Abstract**

Microwave induced resistance oscillations (MIROs) were studied experimentally over a very wide range of frequencies ranging from ~20 GHz up to ~4 THz, and from the quasi-classical regime to the quantum Hall effect regime. At low frequencies regular MIROs were observed, with a periodicity determined by the ratio of the microwave to cyclotron frequencies. For frequencies below 150 GHz the magnetic field dependence of MIROs waveform is well described by a simplified version of an existing theoretical model, where the damping is controlled by the width of the Landau levels. In the THz frequency range MIROs vanish and only pronounced resistance changes are observed at the cyclotron resonance. The evolution of MIROs with frequency are presented and discussed.

---

## 1. Introduction

Microwave induced resistance oscillations (MIROs) observed on very high mobility samples have attracted considerable interest, partly because of the existence of zero-resistance states under certain conditions[1,2,3] Several theoretical models were suggested to explain this phenomenon.[4,5,6,7] It was recently shown [8] that a simplified version of the model based on spatially indirect inter-Landau level (LL) transitions[4,5] describes the MIROs waveform accurately in the quasi-classical regime of large filling factors. It should be noticed that other theoretical approaches based on a non-equilibrium distribution function[6] or the quasi-classical electron orbit dynamics[7] produced similar results in regards to the waveform and phase of the oscillations. On the other hand, neither of the existing theories can explain other MIROs characteristics, such as: (i) the absence of dependence of the waveform on the right/left circular polarization[9], (ii) the absence of dependence on the LL index change $\Delta n$ for the probabilities of the inter-Landau level transitions contributing to the MIROs[8], and (iii) the very high frequency dependence of the MIROs, in particular, where these oscillations have never been observed at optical or sub-millimeter frequencies.

In this work we experimentally investigate the frequency dependence of the MIROs and their evolutions over a very wide range of frequencies, ranging from ~20 GHz up to ~4 THz, starting with the quasi-classical regime and up the quantum Hall regime. We show that MIROs start to deviate from the theoretical model at frequencies above ~120 GHz and vanish at higher frequencies above ~200 GHz.

## 2. Experimental results

Experiments were performed on two samples of GaAs/AlGaAs hetero-structures with a high mobility two dimensional electron gas (2DEG) confined at the interface. After a brief illumination with a red LED, the 2DEG in the first sample attained mobility of $4\times10^6$ cm$^2$/Vs and in the second sample $8\times10^6$ cm$^2$/Vs at a temperature of 2 K. The electron concentration was around $2.0\times10^{11}$cm$^{-2}$ in both samples. The samples were placed in a $^4$He cryostat equipped with a superconducting solenoid. Three sources of microwave radiation were employed depending on the frequency range. In low-frequency experiments between 20 and 50 GHz an Anritsu signal synthesizer (model 69377B) was used and the microwaves were delivered to the sample by means of a semi-rigid coaxial cable equipped with a small antenna at the end.[8] In the medium frequency range between 80 and 220 GHz a tunable klystron microwave generator (model ΓC-03) in tandem with a frequency doubler was used.
For higher frequencies in the tera-Hertz range a far-infrared gas laser pumped with a $CO_2$ laser was employed. In these last two cases the MW radiation was delivered into the cryostat using an oversize thin-wall stainless steal pipe. For each set of experimental traces the MW power at the sample was maintained at a constant level which was controlled by a carbon thermo sensor for all frequencies.[10]

Typical experimental traces of the MIROs on the first sample ($\mu\approx4\times10^6$ cm$^2$/Vs) at low frequencies are shown in Fig. 1 (a) and the corresponding theoretical curves are presented in Fig. 1(b) which were calculated using the following equation:[8]

$$\Delta J_x(B) \propto \mathrm{A}\int d\varepsilon\ \left[n_F(\varepsilon)-n_F(\varepsilon+\omega)\right]\ \nu(\omega)\frac{\partial}{\partial\varepsilon}\left[\nu(\varepsilon+\omega)\right]$$

where $$\nu(E,B)=\sum_{i=0}^{\infty}\frac{eB}{\pi\hbar}\frac{1}{\pi\Gamma\left(1+(E-E_i)^2/\Gamma^2\right)} \qquad (1)$$

is the electron density of states in a quantizing magnetic field, and Γ is the width of the Landau levels. It is evident from the figure that in this frequency range the theory fits the data very well, where we used the same value of Γ=28 μeV for all curves.

Figure 2 shows experimental (a) and simulated (b) traces of the MIROs on sample 2 ($\mu\approx8\times10^6$ cm$^2$/Vs) in the medium frequency range from 80 to 226 GHz. As we can see from the figure, at frequencies higher than ~150 GHz the MIROs progressively become smaller and completely vanish at frequencies higher than 230 GHz that sets the upper frequency limit for the observation of the MIROs. It is clear that the theoretical model fails to describe this behavior.

Figure 3 shows the microwave induced changes in the resistance in the THz range along with the magnetoresistance trace for the Shubnikov-de Haas oscillations $R_{xx}$. In this frequency range, pronounced resistance changes are observed in the form of relatively sharp peaks under the condition of a cyclotron resonance at $\omega=\omega_c$ but no evidence for MIROs is seen. The observed resistance changes can be qualitatively understood in terms of the bolometric-type response of the 2DEG resistance.

Let us examine what happens with the fitting parameter $\Gamma$ by following the evolution of MIROs from low to high frequencies. Figure 4 shows the plot of the LL width $\Gamma$ as a function of the MW frequency obtained by fitting of the data in Fig. 2 with eq.(1). Since the fitting was done in approximately the same magnetic field range from ~0.02 to 0.4 T, the changes in $\Gamma$ cannot be due to a possible dependence of $\Gamma$ on magnetic field. Indeed, as seen from the plot, $\Gamma$ remains constant for frequencies below 120 GHz, as mentioned above, where the model works. Between 120 and 156 GHz the LL width $\Gamma$ increases stepwise from 25 to 60 μeV and at 226 GHz it increases even further and MIROs vanish at higher frequencies. Such behaviour is not addressed in existing theoretical models and has to be understood. Qualitatively, the damping of MIROs (equivalently, effective broadening of the LL width) can be explained if an additional coulping is involved in the interaction between light and the electrons. Most likely, this additional coupling is related to magneto-plasmon excitations in a finite size sample due to the strong electrodynamic interaction between radiation and mobile charges.[8,11] If the lifetime of the plasmon excitation depends on the MW frequency, which would lead to the experimentally observed broadening of the total effective width $\Gamma_{tot}=\Gamma+\Gamma_p$. This additional coupling might also be at the source for the lack of dependence of MIROs on circular polarization[9] and the transition probability insensitivity to the LL index change $\Delta n$.[8]

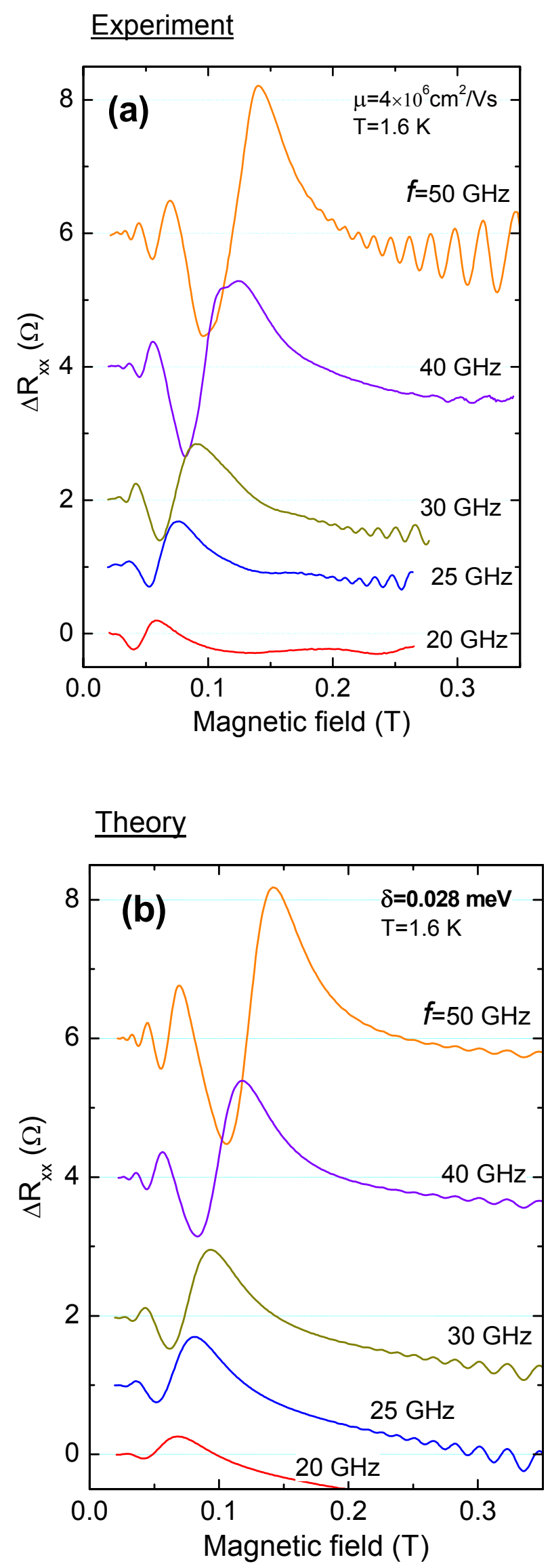

Fig. 1. MIROs traces for different MW frequencies from 20 to 50 GHz on GaAs/AlGas sample 1 ($\mu\approx4\times10^6$ cm$^2$/Vs); (a) experiment, (b) calculated with eq. (1) using the same Landau level width $\Gamma$=28 μeV. Traces are shifted vertically for clarity.

## 3. Conclusion

We studied the evolution of the MIROs over a wide range of frequencies. At MW frequencies below 150 GHz the MIROs are well described by an existing theoretical model, where the only fitting parameter is the LL width $\Gamma$, which remains constant for f<120 GHz but increases rapidly for higher frequencies. At even higher frequencies the MIROs disappear and only the cyclotron resonance is observed.

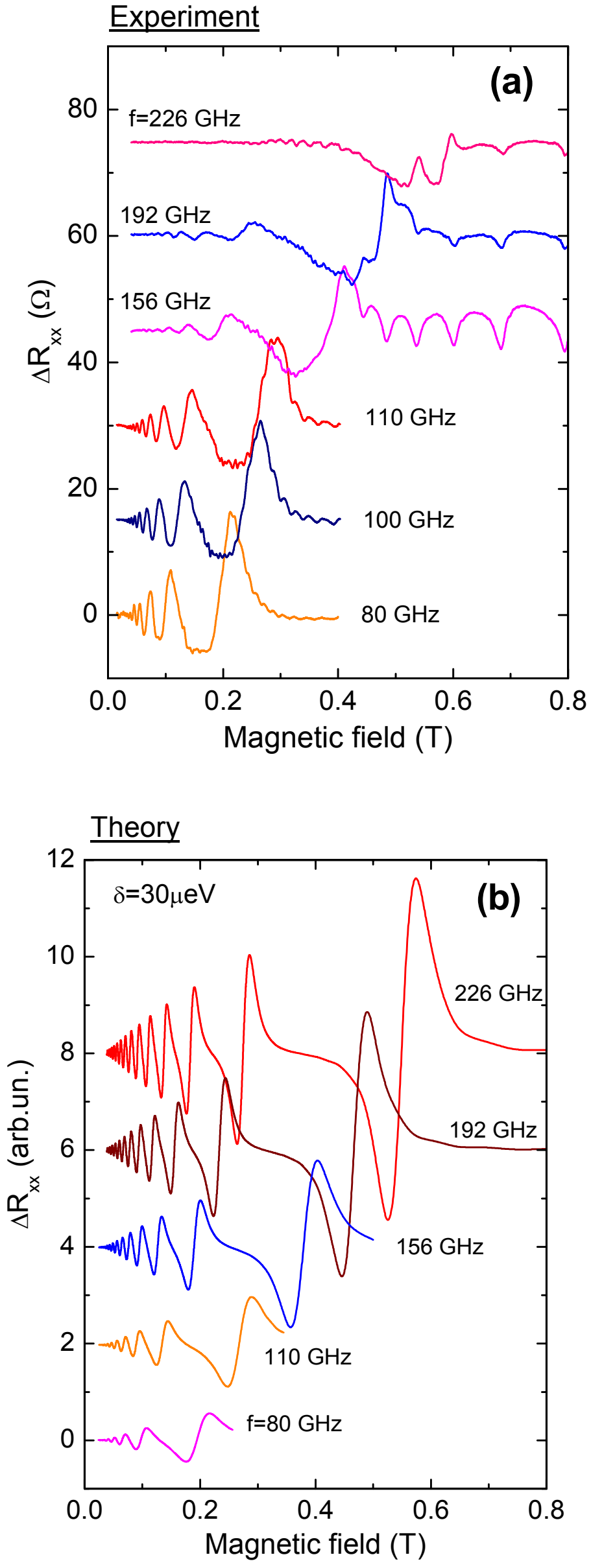

Fig. 2. MIROs ($\Delta R_{xx}=R_{xx}$(MWs on)-$R_{xx}$(no MWs)) for different MW frequencies from 80 to 226 GHz using GaAs/AlGas sample 2 ($\mu\approx 8\times 10^6$ cm$^2$/Vs); (a) experiment, (b) theory eq. (1) with the same Landau level width Γ=30 μeV.
Traces are shifted vertically for clarity.

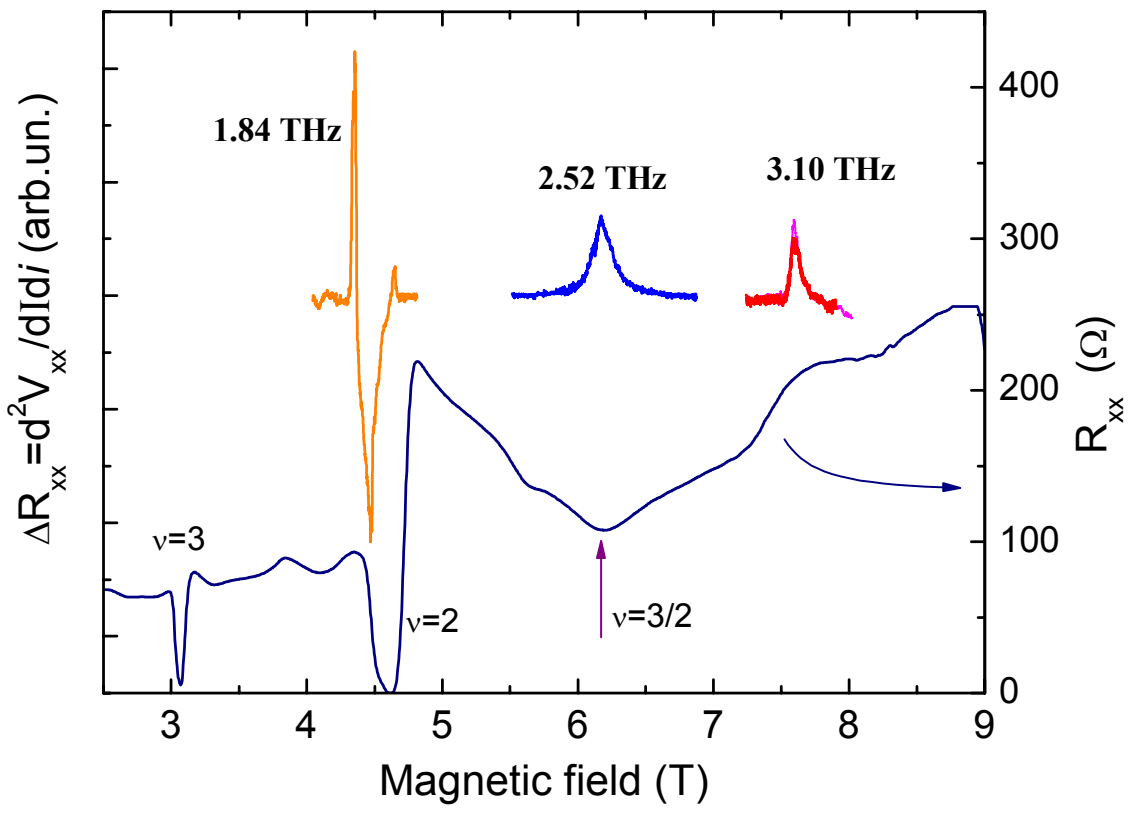

Fig. 3. Photo-response of the high mobility 2DEG in the quantum Hall regime at THz frequencies.

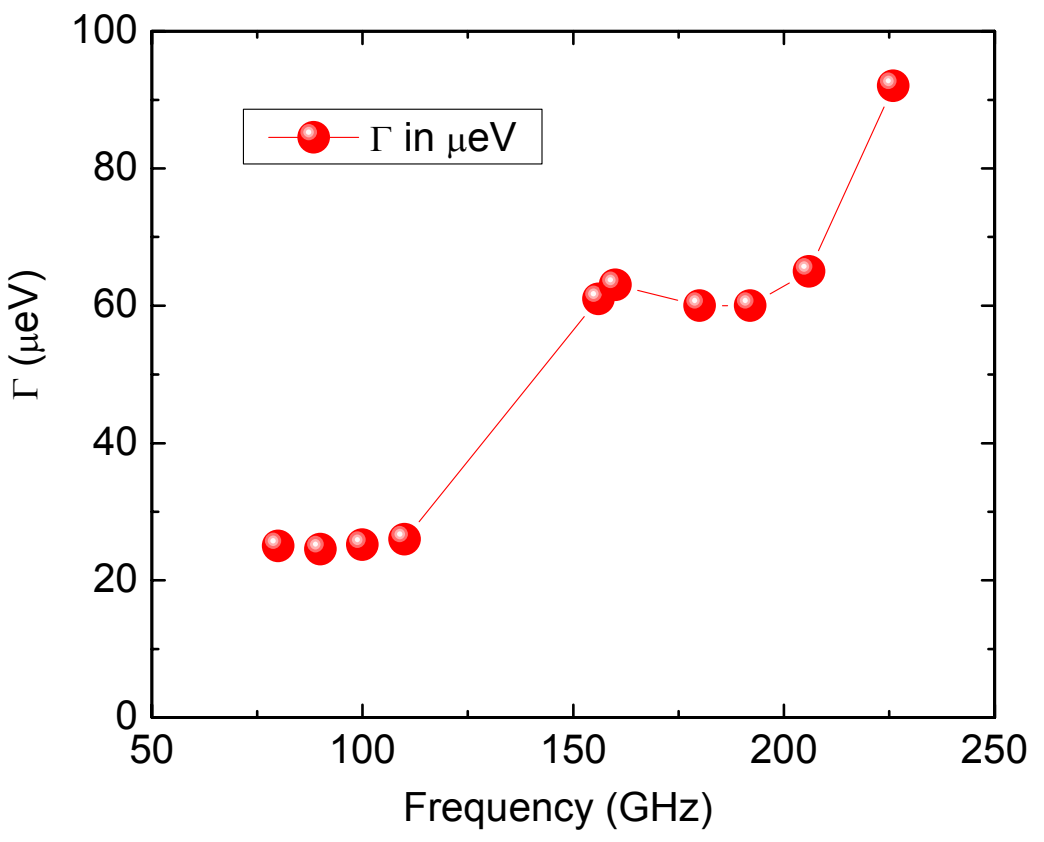

Fig. 4. Landau level width determined from MIROs and fitted by eq. (1).